# Spectrally-resolved photodynamics of individual emitters in large-area monolayers of hexagonal-boron nitride

H. L. Stern*[1] and R. Wang[2], Y. Fan[2], R, Mizuta[2], J.C. Stewart[2], L.M. Needham[1], T. D, Roberts[3], R. Wai[3], N. S, Ginsberg[3], D. Klenerman[1], S. Hofmann*[2] and S. F. Lee*[1.]

[1] Department of Chemistry, University of Cambridge, Lensfield Road, CB2 1EW, Cambridge, United Kingdom.

[2] Department of Engineering, University of Cambridge, JJ Thompson Avenue, CB3 0FA, Cambridge, United Kingdom.

[3] Department of Chemistry, University of California, Berkeley, CA 94720.

* Corresponding authors

**Abstract**

Hexagonal boron nitride (h-BN) is a 2D, wide band-gap semiconductor that has recently been shown to display bright room-temperature emission in the visible region, sparking immense interest in the material for use in quantum applications. In this work, we study highly crystalline, single atomic layers of chemical vapour deposition (CVD)-grown hexagonal boron nitride and find predominantly one type of emissive state. Using a multidimensional super-resolution fluorescence microscopy technique we simultaneously measure spatial position, intensity and spectral properties of the emitters, as they are exposed to continuous wave illumination over minutes. As well as low emitter heterogeneity, we observe inhomogeneous broadening of emitter linewidths and power law dependency in fluorescence intermittency, this is in striking similarity to previous work on quantum dots. These results show that high control over h-BN growth and treatment can produce a narrow distribution of emitter type, and that surface interactions heavily influence the photodynamics. Furthermore, we highlight the utility of spectrally-resolved wide-field microscopy in the study of optically-active excitons in atomically thin two-dimensional materials.

**Introduction**

Two-dimensional materials are rapidly emerging as promising for use in the areas of electronics,[1] materials science,[2] photovoltaic, light-emitting devices,[3] and quantum emission.[4,5,6] In the case of quantum technologies, the ideal solid-state system is sought that displays atomic-like states, which can be deterministically prepared and read-out optically.[5,6]

Hexagonal-boron nitride (h-BN) has recently joined the small family of materials that display single photon emission.[7] Previously, h-BN was widely used as a dielectric layer in graphene devices[8,9] and was recently shown to display bright, visible single photon emission at room temperature.[7] The emission has been compared to that of nitrogen-vacancy (NV) centres in diamond, where the excitonic defect is due to an atomic-scale point defect.[10,11] The emissive defect state in h-BN is predicted to be formed by either a boron vacancy ($V_B$), nitrogen vacancy ($V_N$) or an antisite complex ($N_BV_N$), in which a boron atom is replaced by a nitrogen atom with an absent neighbouring nitrogen atom.[7] Consequently, the past few years have seen work on single-photon emitters in h-BN accelerate, but progress towards device integration has been hindered by the high variation in the frequency of the emission.[12,13,14]

An explanation for the wide-range of emission energies reported is the use of heterogeneous h-BN samples. So far, the vast majority of optical studies of h-BN have been carried out on multilayer h-BN samples without a well-defined sample thickness or morphology, namely liquid[7,12,13,] or mechanically exfoliated[15,16,17] h-BN. Due largely to the difficulty in the sample preparation, studying h-BN produced from scalable, integrated growth methods, such as CVD, has not been a priority for optical studies. However, the most promising method to produce h-BN with well-defined thickness over large area is chemical vapour deposition (CVD). [18,19] While layer thickness is known to strongly affect the band structure of h-BN,[20] the optical properties of large crystalline domain monolayer h-BN are yet to be fully explored.

Here we focus on large areas (~100 mm$^2$) of continuous, single crystalline (with domain sizes exceeding 2500 μm$^2$) CVD-grown h-BN monolayers.[18] These monolayers represent a homogenous surface (without large variation in thickness, edges or high density of grain boundaries) and thus present a well-defined model system to study defect emission in h-BN. The h-BN films are grown on a platinum catalyst and directly transferred using mechanical delamination onto glass substrates for the optical measurements.[18]

It is typically difficult to find techniques that can locate and non-destructively probe such thin and optically transparent samples. Established structural techniques, such as high resolution TEM[21] and scanning tunnelling microscopy (STM),[22] address small areas at a time and can induce new defects to the sample.[23] Traditional optical techniques, such as confocal microscopy, typically do not enable uniform excitation fluence across a wide area.

To study this material we use a custom-built spectrally resolved single-molecule fluorescence microscope, previously used in a biological context for multidimensional super resolution imaging,[24] that enables high-throughput measurements of relatively large areas (>10 μm$^2$) and simultaneous collection of spatial and spectral information. Surprisingly, we find that these h-BN monolayers produce emitters with high uniformity in emission wavelength. Unlike previous reports,[13] we find one type of emitter with emission predominantly at 580 nm ± 10 nm. We also did not treat the h-BN with high temperature argon annealing or electron irradiation to induce or stabilise emitters.[13] We observe that the emitters display fluorescence intermittency

('blinking') with 'on' and 'off' times described well by a power-law distribution, similar to that previously shown in semiconductor nanoparticles or quantum dots.[25,26,27] From an analysis of time-resolved spectral changes in conjunction with photodynamics, we find that single emitter spectral diffusion is associated with blinking, leading us to conclude that a control of h-BN atmosphere or surface chemical treatment may improve spectral properties.

**Results and Discussion**

**Power-law blinking of surface emitters in h-BN.**

The h-BN monolayer was grown following protocol explained previously in detail.[18] In Figure 1(a) the Raman spectra for a monolayer h-BN on $SiO_2$ is shown (red trace). A Raman peak at 1369 $cm^{-1}$ is seen, and is not observed in the blank (black trace), indicating the presence of monolayer h-BN.[18] In addition, transmission electron microscopy (TEM) revealed the h-BN monolayer is made up of large (>10 $\mu m^2$) aligned domains (Figure S1). Combined, these results support that our preparation method created highly crystalline monolayer h-BN.

Wide-field super-resolution microscopy is used to illuminate large areas of the h-BN monolayers. Continuous-wave laser illumination is used to confine the excitation geometry (~200 nm axially) and excite the sample via an evanescent wave formed via total internal reflection (TIR) at the glass-air interface (Figure 1(d)).[28] This geometry enables high sensitivity (signal/background ratio) typically used in single molecule microscopy. The Stokes-shifted emitted light from the sample is collected via the same objective and imaged onto a (single-photon sensitive) electron multiplied charge coupled device (EMCCD) camera. Images are stored as successive TIFF stacks with an exposure time of 100 ms (see supplementary information for details).

Figure 1(b) shows a TIR fluorescence image of monolayer h-BN when excited with continuous-wave 532 nm laser light, in ambient air at room temperature. 532 nm (2.33 eV) is well below the band-gap of h-BN (~6 eV),[29] therefore does not excite the bulk of the material, but does interact with the sub band-gap defect states.[7,30,31] Each diffraction-limited fluorescence puncta or point spread function (PSF) represents a single emitter. It can be both analysed in time (as shown in time montage in Figure 1(c)) and can be fit to a 2D Gaussian to super-localise its spatial position (typically 10-100 nm), where the localisation precision is related to the fit error of the PSF and the number of photons (N) emitted: $\sigma_{x,y} \sim \frac{\sigma_{PSF}}{\sqrt{N}}$. For the emitters in h-BN, we obtain typical localisation precisions ranging from 12 nm - 30 nm (Figure S5). We observe intermittency in the emission (Figure 1 (c)), behaviour that is typically attributed to a molecule or single quantum dot and representative of a single emitter, not an ensemble.[25]

To characterise the fluorescence intermittency observed, in Figure 2(a) we plot the emission from a single emitter in monolayer h-BN over 100 s. The intensity of light from the emitter fluctuates over time. Using a threshold (details in Methods) to distinguish between 'on' and 'off' states, 'on' and 'off' times are calculated.

Figures 2 (b) and (c) show the 'on' and 'off' time distributions obtained from histograms of 'on' and 'off' time, for 208 emitters measured across a total surface area of 0.018 mm$^2$ of three h-BN monolayer samples. Figures 2(b) and (c) represents over 2.0 x 10$^6$ and 13,000 data points respectively. The probability distribution for both times can be readily fit to a power law of the form $P(t) \sim t^{m^{on/off}}$ where m$_{ON}$ = 2.31 and m$_{OFF}$= 1.71. With a bin size of 100 ms, we find that the 'off' times for the h-BN emitters are spread over three orders of magnitude in time (10$^1$-10$^4$ ms), whereas the 'on' times show slightly less variation ($10^1$-$10^3$ ms) and the gradient to the power law fit is at the higher end for what has been observed for other nanoscale systems (1.7-2.2).[25,32] Interestingly, such power law blinking behaviour has been found and well studied in other nano- optical systems, such as quantum dots, GaN single photon emitters,[33] nanodiamonds[34] and single molecules,[35] where it indicates either population of a 'trap' or formation of a fluorescent inactive ('dark') triplet state. We find here that the 'off' times show little dependence on laser intensity (inset Figure 2(c)), whereas the 'on' times show a weak dependence, which is the same trend seen in quantum dots.[27] No intensity dependence for the 'off' times suggests the recovery mechanism of the emissive exciton is not light induced.[27] A power law distribution of 'off' times, as opposed to an exponential fit, points to a number of potential physical mechanisms for the formation of the non-emissive state, such as an exponential distribution of trap depths, fluctuating non-radiative decay rates or a distribution of tunnelling distances between photo-induced exciton and trap state.[25,26,27] Characterisation of the emitter photophysics may help us to understand the photoinduced dynamic processes occurring on the surface of h-BN in an ambient atmosphere.

Fluctuating emission intensities have been observed from h-BN previously[36,37,38,39] but the cause of the instability has not been rigorously explored. Evidence points to higher emitter stability in thick, multilayer h-BN,[7] indicating that the blinking behaviour could be induced by proximity to the surface. Surface traps may be formed via impurities or surface adsorbed molecules, which have been shown to chemically react with h-BN defects.[40] As such, h-BN quality: growth, treatment and transfer, are likely to be critical parameters in controlling how surface photoinduced emitters behave.

**Spectrally-resolved wide field imaging reveals mono-colour emitters.**

While blinking of single emitters is disadvantageous for quantum applications, it enables us to understand the photodynamics of individual surface-bound defects. Blinking dynamics have been a critical tool in understanding trapping behaviour, and designing methods of mitigating it, on the single particle level for quantum dots.[27,41] With a technique that has been deployed in research on quantum dots before,[26,42,43] we perform spectrally-resolved measurements to determine the nature of the emitters.

Using a bespoke instrument we have previously reported,[24] we simultaneously measure emitter position, blinking trajectories and spectra. This is physically implemented on the set-up described above, via the addition of a blazed diffraction grating, placed in the optical emission path before the image plane (Figure 3(a)). The two major diffraction orders (zeroth and first order) are projected onto different spatial regions of the EMCCD detector and recorded simultaneously (Figure 3 (a) and (b)). The zeroth order transmission corresponds to the spatial position of emitters on the surface of the h-BN (Figure 3(d)) and is used to super-localise emitter position in x,y. The first order diffracted light is the spectral information of the fluorescent signal (Figure 3(c)). The spectral resolution is related to the localisation precision of the spatial image of the emitter (analogous to the slit width in a conventional spectrometer), thus related to the detected photons above background. We analyse all emitters with a fitted spatial resolution of < 30 nm, corresponding to integrated photon counts of ~1.0 x $10^3$ photons/frame and a spectral precision in the range of 2-8 nm, with a typical model value of 4 nm (Figure S4). The pixel-to-wavelength calibration of the detector is performed using fluorescent beads and takes account of spectral dispersion, as described in reference 24 and in Figure S2.

A typical emission spectrum obtained for an emitter in monolayer h-BN using this method is shown in Figure 3(e). This spectrum is the integrated emission spectrum taken from a Z-stack of 1000 frames. The spectrum is broad, asymmetric, with a peak, or zero phonon line (ZPL), at ~550 nm, a breadth of ~50 nm (~5 x spectral resolution) and low energy sideband, consistent with the emission observed previously for emissive defects in h-BN.[7,12,44] As previously noted,[7,12] the line-width of emission of defects in monolayers is greater than the line-width observed for defects embedded in multilayer h-BN.[45] Due to the similarity in shape and ZPL position between spectrum in Figure 3(e) and the reported spectra of single photon emitting defects in multilayer h-BN,[12,13,14,15] we predict that the emission from the CVD-grown monolayer is likely to derive from defects of a similar structure. A control experiment of a blank glass slide subjected to the same polymer transfer confirmed that the emitters show a distinct emission profile from organic contaminant (emission centred at 610-620 nm) (Figure S8) and that organic contaminant bleaches within the first 20-50 s (Figure S7).

This inhomogeneous broadening is analogous to that observed in atomic or molecular transitions where radiating atoms or molecules interact differently with the environment, leading to a distribution of transition energies in the ensemble emission spectrum.[46] The FWHM of monolayer emission observed here (Figure 3(e)) suggests the transition energy can be modulated up to ~190 meV, consistent with reports of the energy splitting observed between the ZPL and phonon sideband for h-BN defects.[7,12,15,38] Interestingly, we find that despite the inhomogeneous broadening of single-defect line-widths, the ensemble population of emitters is more *homogenous* than observed in thicker h-BN.[12]

For every emission spectrum we resolve we fit the peak with a Gaussian and obtain the peak fluorescence emission wavelength. In Figure 3(f) we present a histogram of the distribution of peak emission wavelengths from >8,000 emitter emission spectra, including multiple spectra from a single emitter, collected from h-BN monolayers from two independent growth runs. The bin size for the histogram is 10 nm, slightly larger than the spectral resolution for the weakest emitters (8 nm). The data is not filtered to exclude any emission detected at 610 - 620 nm, a region we have identified as potentially due to contamination (Figure S8). We observe a preference of emission maxima at 580 ± 10 nm, a region repeatedly identified as a dominant ZPL for defects in multilayer h-BN,[12,15,16] consistent with previous results from CVD-grown monolayers.[40] There is a distribution of emitters with emission maxima at higher energies- between 550 nm and 580 nm- but fewer emitters with red-shifted emission maxima of 600 nm. Others have reported the presence of two distinct spectral classes of quantum emitters in h-BN – those that emit at ~570 nm and those that emit lower than 650 nm,[13] Here, we find no evidence of this lower energy emitter, which we attribute to the different growth and processing of the h-BN samples. We see any preference for emitters at edges or grain boundaries. Emitters appear to be randomly distributed across the monolayer.

The emitters on monolayer h-BN show a high degree of reproducibility when we compare individual growth runs, for growth 1 (67 % of emission spectra are at 580 nm ± 10 nm, for example Figures 3(f) inset). We predict the difference in emission maxima between growth runs to be due to the varied degree of strain imparted into the monolayer during growth and/or transfer.[16,47]

These findings suggest the monolayers are made up of predominantly one type of emitter that can interact with its environment to give a broadened line-width. To understand more about this environmental interaction we analysed the spectrally-resolved photodynamics. Across the h-BN samples, we observe that emitter behaviour ranges from well-defined, telegraph-like on/off transitions (as displayed in Figure 2(a)) to emitters that transition between continuous distributions of emission intensities. To illustrate the range of behaviour observed, in Figures 4(b) and (g) we show luminescence counts against time for two different emitters, emitter 1 and emitter 2. The fluorescence intensity trajectory for emitter 1 (Figure 4(b)) shows a well defined 'on' and 'off' levels in photon counts that clearly shows two clear intensity distributions (Figure 4(e)). The emitter is dark for the first 30 s before emitting, then subsequently enters one period of darkness before returning to the emissive state for the entirety of the measurement. Correspondingly, the colour of emission is relatively constant (Figure 4(a)), with peak emission energy at 545 nm and FWHM of ~10 nm, displayed in the single emission spectrum in Figure 4(c) and the distribution of all emission maxima in Figure 4(d).

In contrast, Figure 4(g) is the fluorescence trajectory for a second emitter that shows rapid blinking, a well defined 'off' photon count level and a continuous distribution of photon emission events at higher photon counts (Figure 4(k)). The emission maxima vary more than emitter 1 (Figure 4(h) and (i) show emission spectra at different frames in the trajectory) and are on average lower in energy (centred at 560 nm) (Figures 4(j)).

We find that the emitters across the samples display a range of behaviour between these two examples, with no notable correlation with spatial position (previously observed in quantum dots).[42] In Figure 4(l) we plot the distribution of average number of on-off-on fluorescence intermittency cycles ('blinks') for each emitter from one growth run, over 200s of illumination. Unlike Figures 2b and c, each data entry here represents one emitter. As we can see, most emitters (87%) display less than 100 blinks. There is a small population (<1 %) that display a very large number of blinks (>600). We assess the global photobleaching rate of the emitters and find that the average emitter total photon count has reduced by 60% after 200 s (Figure S7).

As our technique enables us to probe the behaviour of individual defect emitters, we analyse the extremes of the population and can selectively interrogate both the very low and the very high blinking emitters, to assess the effect of blinking on spectral properties. In Figure 4(m) we plot all spectral maxima collected for emitters that show an average number of blinks greater than 100 and a median 'off' time larger than 30 s, the 'low' blinking emitters, and in Figure 4(n) we plot the same data for emitters that blink over 600 times and have a median 'off' time of less than 30 s. This separation selects for emitters that are constantly blinking (~1% of total population), and those that exist in one state for long periods of time (~10 % of total population).

We observe that emitters that show low blinking have a narrower distribution of emission peaks (Figure 4(n)) while emitters that blink more show slightly more broadened emission (Figure 4 (m)).

Combined, these results show a correlation between blinking and spectral diffusion for single emitters in monolayer h-BN, closely reminiscent of the behaviour observed in quantum dots, where perturbations in the local electric field on the surface of the quantum dots, or systematic surface oxidation,[48] manifests as shifts in emission frequency.[49] More generally, the demonstration of power law blinking, and the overlap with quantum dot systems, is significant; work over decades for quantum dots has used single quantum dot fluorescence intermittency to understand the complex photophysical phenomena at play and then exploiting it to create rational devices. For example: the origin of blinking,[25,26,27] discovery of photoluminescence enhancement[50] and spectral bluing,[23] chemical and passivation methods for reducing dot blinking[51] and improving quantum yield,[52] have all been important in the development of quantum dots for different biological[53] and energy applications.[54] It may be possible to draw on this literature to enhance h-BN emitter properties. Certain approaches to control blinking, such as use of encapsulation or controlled atmospheres that prevent oxidation, may enable single layer 2D materials to be viable candidates for future optical applications.

**Conclusion**

We have shown that wide-field microscopies that are routinely used in biology can provide high-resolution, high-throughput and multidimensional approaches to study the most challenging atomically-thin material systems.

In the search for deterministic, narrow-band, stable single photon emission, these results show that h-BN growth is important to consider. We show that CVD-grown monolayers with large crystalline domains represent homogenous samples and produce predominantly one type of emitter, with highly reproducible emission maxima at 580 nm ±10 nm. The emitter line shape is consistent with defects measured in thicker h-BN and we propose they represent defects of the same structure. Our results show atomically-thin h-BN can possess these emitters, without the need for high temperature argon annealing or ion irradiation, making this an attractive material for future use.

However, we show that emitters exposed to the surface suffer from disorder-inducing interactions, most likely with surface charges. Future work should involve control of surface species, via encapsulation for example. Methods of controlling surface contamination, without compromising optical extraction efficiency, will make atomically thin h-BN an attractive candidate for future optical quantum applications.

## Materials and Methods

**Sample Preparation**

h-BN was grown via a CVD-method previously reported.[18] In brief, h-BN is grown on a Pt catalyst by exposure at temperatures of 1000 °C to borazine. Continuous monolayers of h-BN form after a short period of growth. The samples are transferred onto the glass substrate using mechanical delamination. Hereby, a stamp is applied, and the h-BN layer can be directly peeled off the catalyst due to the weak interaction between the two. The stack consisting of stamp and h-BN are transferred onto a glass substrate and the stamp is removed. Prior to measurement all samples were annealed for 3 hours at 500 °C in ambient atmosphere to remove all traces of organic contaminant from the transfer process.

**Structural Techniques**

Raman measurements were performed with a Reinshaw inVia confocal Raman microscope. The TEM is carried out on a FEI Tecnai Osiris S/TEM under 80keV. The diffraction mapping is done by taking a diffraction pattern at grid points with step size around 1.8 $\mu$m in both the x and y directions. A selective aperture is used when taken the diffraction so that the area of measuring is with about 1.4 $\mu$m radius. The x and y position is recorded by the reading from the piezo-stage.

**Optical set-up**

Fluorescence imaging was performed using a homebuilt, bespoke inverted microscope (Olympus IX73) coupled to an electron multiplied charged coupled device (EMCCD) camera (Evolve II 512, Photometrics, Tuscon, AZ). The microscope was configured to operate in objective-type total internal reflection fluorescence (TIRF) mode. A 100 mW 532 nm continuous wave diode-pumped solid-state laser

(LASOS Lasertechnik GmbH, Germany) was directed off a dichroic mirror (Di02-R532-25x36, Semrock) through a high numerical aperture, oil- immersion objective lens (Plan Apochromat 60 x NA 1.49, Olympus APON 60XOTIRF, Japan) to the sample coverslip. Total internal reflection was achieved by focussing the laser at the back focal plane of the objective, off axis, such that the emergent beam at the sample interface was near-collimated and incident at an angle greater than the critical angle for a glass/air interface for TIRF imaging. This generated a ~ 50 $\mu$m diameter excitation footprint with power densities in the range ~100 mWcm$^{-2}$ at the coverslip. The emitted fluorescence was collected through the same objective and further filtered using a longpass filter BLP01-532-25 (Semrock, USA) and a bandpass filter FF01-650/200-25 (Semrock, USA) before being expanded by a 2.5 x relay lens (Olympus PE 2.5 x 125). Finally, a physical aperture and transmission grating (300 groves/mm 8.6° Blaze angle –GT13-03, Thorlabs) were mounted on the camera port path before the detector. The camera-to-grating distance was optimised using Tetraspeck beads (0.1 $\mu$m, T7297, Invitrogen) such that undiffracted and first order diffraction was visible on the same image frame the fluorescence image was projected onto the EMCCD running in frame transfer mode at 20 Hz, with an electron multiplication gain of 250, operating at -70°C with a pixel size of 16 $\mu$m and automated using open source microscopy platform Micromanager.

Images were acquired at a frame rate typically 100 ms and 1,000-2,000 frames were acquired. Each pixel on the image was equal to 107 nm. The co-ordinates corresponding to the localizations within the spatial part of the images were determined using the peak fit plugin for ImageJ.

Samples were measured in air, with the h-BN facing upwards, illuminated from the other side of the glass.

**Spectral Calibration**

The location of the spectral domain on the EMCCD chip is strongly dependent on the orientation and position of the diffraction grating, and so to ensure that the wavelengths determined were correct and therefore comparable, the instrument was calibrated before imaging each day, and following re-alignment.

To calibrate the instrument, TetraSpeck beads were imaged by exciting at 405 nm, 532 nm and 633 nm and the emission collected from 480 to 760 nm (Supplementary Figures 1-4) (100 frames were collected with a frame rate of 50 ms). The positions of the beads within the spatial domain of the image were determined using the PeakFit plugin (an imageJ/Fiji plugin of the GDSC Single Molecule Light Microscopy (SMLM)) package for imageJ using a typical 'signal strength' threshold of ~30. The fit results values are then saved as a text file in the directory containing the image for use the subsequent spectral analysis below.

The spectral component of the image was analysed using a custom written python code. The position of these three peaks allows both for the aberration correction (due to the relationship between these and the spatial positions of the beads), and for the pixel-to-wavelength ratio to be determined. For each localisation (with co ordinates $x$, $y$, frame number), the spectrum for each of the three peaks was averaged over a width of 3 pixels in the $x$-direction (to account for the spectra being wider than one pixel). Each intensity profile was fit to a Gaussian distribution to determine the centre positions of the peaks. Fits that gave negative amplitudes, centres outside of the range, widths <1.5 pixels or >20 pixels were discarded.

Thresholding analysis for 'on' and 'off times was done using the same PeakFit plugin as described above. Emitters were deemed 'on' if they were detected with the following PeakFit parameters: calibration: 107 nm/px, gain: 250, exposure timew: 100 ms, initial stdev0: 2.0, initial stdev1: 2.0, initial angle: 0.0, smoothing: 0.5, smoothing2=3, search width: 3, fit_solver= [Least Squares Estimater (LSE)], fit_function= Circular, fit_criteria=[Least-squared error] significant_digits= 5, coord_delta= 0.0001, lambda= 10, max_iterations=20, fail_limit= 3, include_neighbours, neighbour_height= 0.30, residuals_threshold= 1, duplicate=0.50, shift_factor= 2, signal_strength= 50, width_factor=2, precision=30.

**Supporting Information**

Supporting Information PDF

**Corresponding Authors**

*hs536@cam.ac.uk, sh315@cam.ac.uk, sh591@cam.ac.uk

**Author Contributions**

HLS, RW and SH conceived the project. HLS designed and conducted experiments and analysed the data, with help from RW,LMN and SL. RW and RM grew the h-BN samples and YF took structural measurements. HLS, RW, RM, YF, DK,JCS, TR, RW, NG, SH and SL discussed the results. HLS wrote the paper with contributions from all other authors.

**Funding Sources**

H.L.S acknowledges support from Trinity College, Cambridge. R.W. acknowledges an EPSRC Doctoral Training Award (EP/M506485/1). J.C.S is affiliated with the Cambridge Graphene Centre and funded by the EPSRC Doctoral Training Centre in Graphene Technology, EPSRC grant EP/L016087/1. R.M was supported by EPSRC Cambridge NanoDTC, EP/L015978/1. We thank the Royal Society for a University Research Fellowship to S.F.L (UF120277). S.H acknowledges funding from the European Union's Horizon 2020 research and innovation programme under grant agreement No number 785219.

# Figures

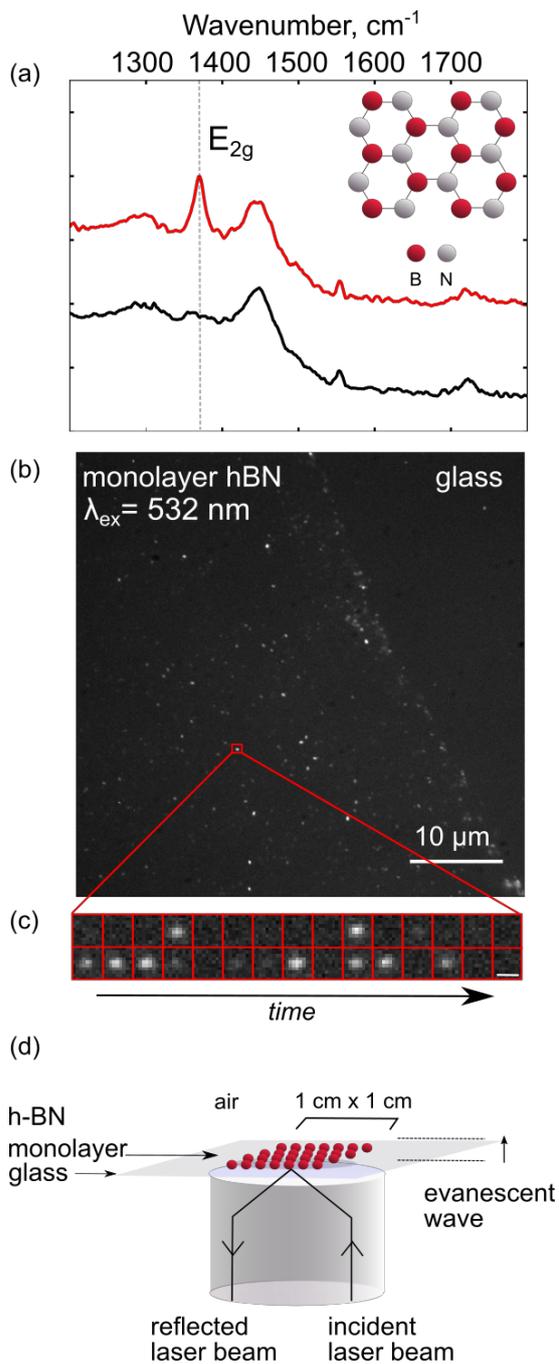

Figure 1. (a) Raman spectra of monolayer h-BN on SiO$_2$ substrate (red) and blank (grey), with monolayer h-BN E$_{2g}$ peak noted with a dashed line. Inset shows a schematic of an h-BN monolayer. (b) A Z-stack of 500 frames from a near-field microscopy image of a monolayer of h-BN on glass, excited with CW laser light at 532 nm. (c) Zoomed images of single 100 ms frames showing the blinking of a single

emitter (highlighted in red in (b). The scale bar is 400 nm. (d) Total internal reflection illumination geometry for measurement of the h-BN monolayer.

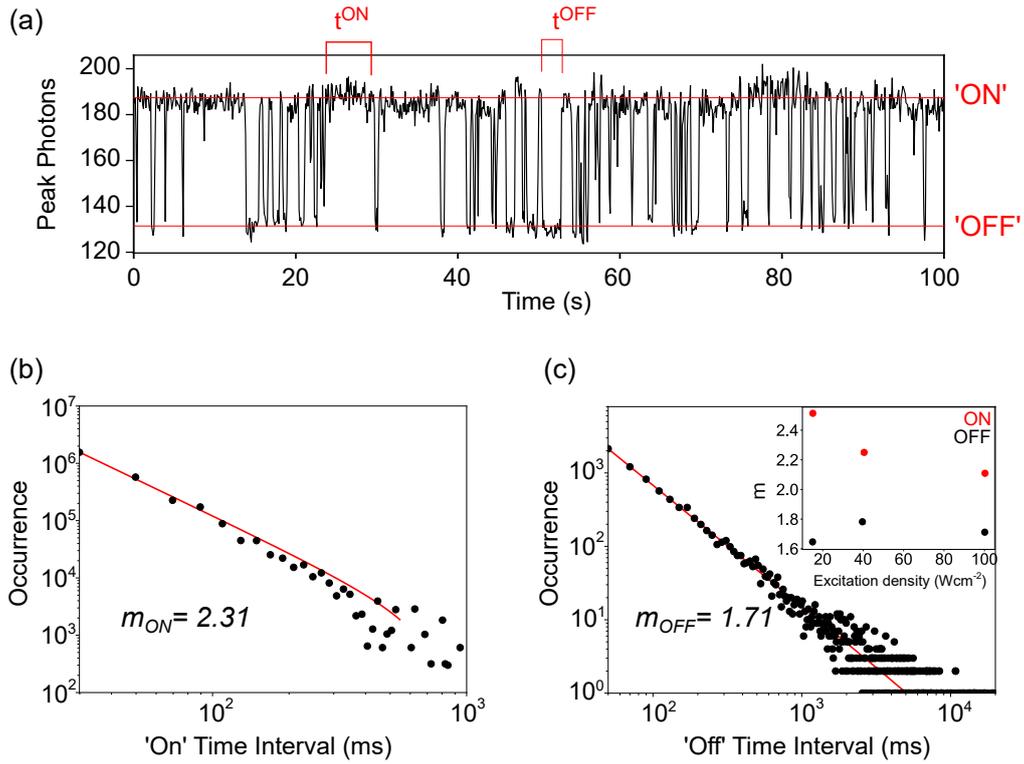

Figure 2: (a) Peak photon counts over time for an emitter in monolayer hBN. Lines 'on' and 'off' are a guide to the eye to represent the photon counts for the emissive and non-emissive thresholds. Probability distribution of 'on' (b) and 'off' (c) times for 208 blinking emitters across 15 measurements with power law fits shown in red. Distributions are composed from histograms of 'on' and 'off' times. Inset shows dependence of gradient on laser excitation energy.

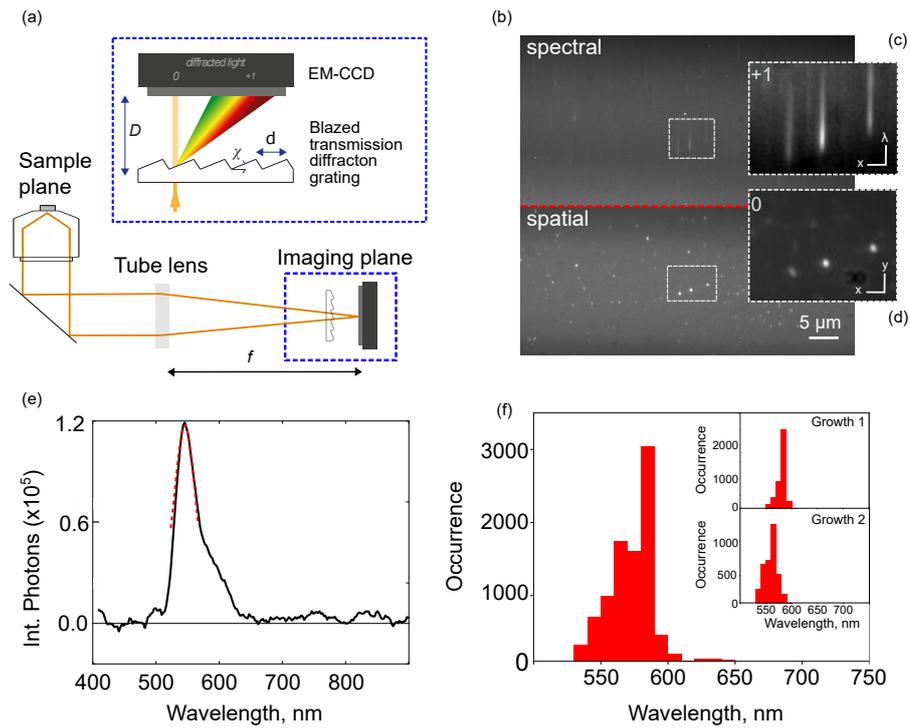

Figure 3. (a) Schematic of the spectrally-resolved measurement geometry. A collimated excitation source is totally internally reflected at the glass-air interface of the sample. Emitted light is directed through a tube lens and towards the diffraction grating and detector. (b) A Z-stack of 1000 frames of the image recorded on the EMCCD. The spatial and spectral domains are contrast adjusted. (c) A zoomed view of the dashed box in the spectral region of image (b). The spectra are obtained from the first order transmission through the diffraction grating. Spatial image of three emissive defects, obtained from the zeroth order transmission through the diffraction grating. The integrated counts obtained for the circled emitter is 5.3 x $10^4$ counts/s. Each axis of the x,y scale bar equates to 1.5 μm. (e) The single emitter spectrum shown in (c), obtained for the defect highlighted in red in (d). Y axis shows integrated photons. The emission maximum is fit with a 2D Gaussian. (f) Histogram showing the peak emission wavelength position across two growths of h-BN monolayers. Inset shows emission maxima obtained for two individual growths.

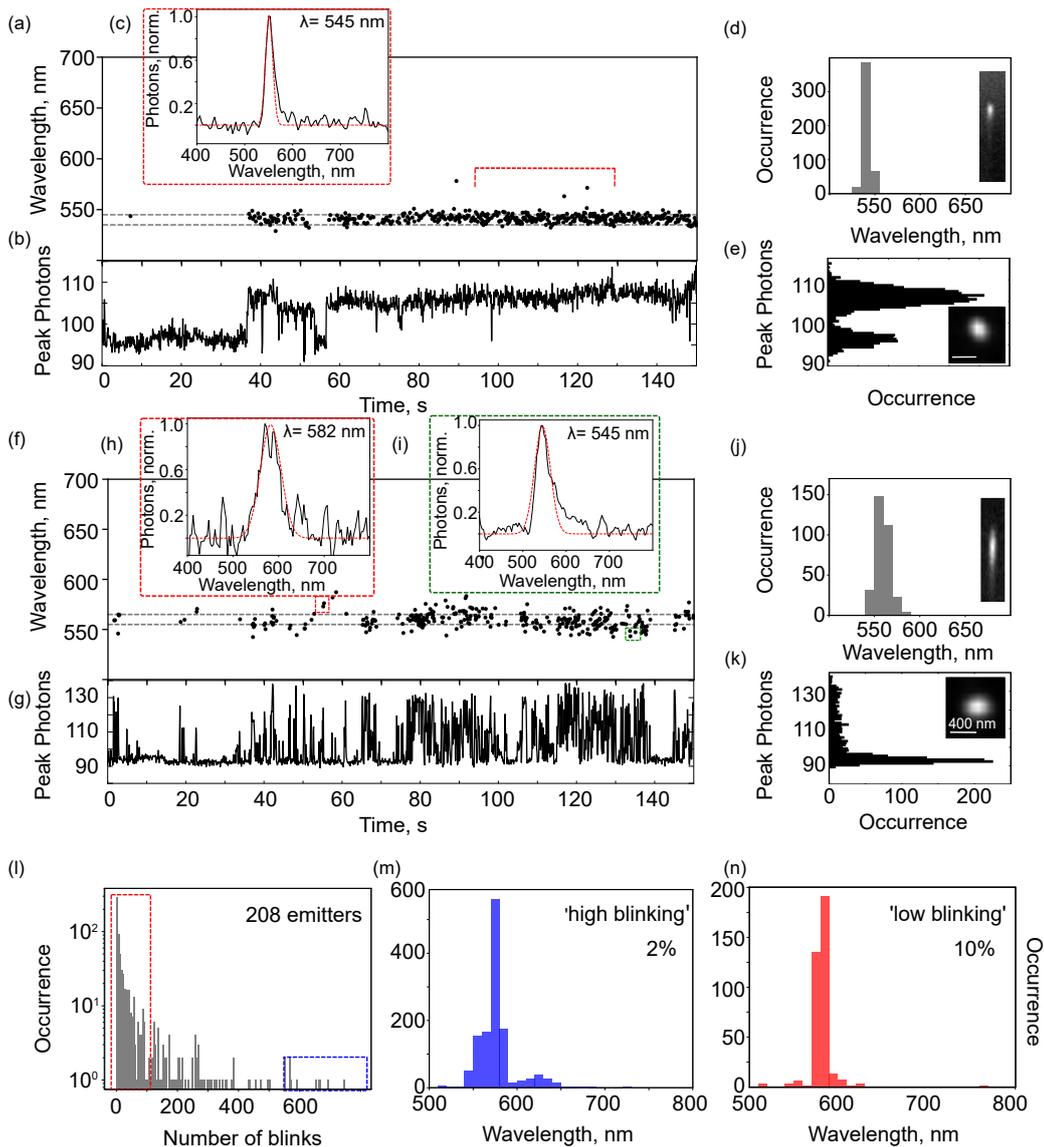

Figure 4. Emission maxima (a) and intensity (b) measured for emitter 1 over 150 s. (c) Integrated emission spectrum for emitter 1 (80-140 s) and histogram of emission maxima (d). Distribution of photon counts (e). Emission maxima (f) and intensity (g) measured for emitter 2 over 150 s. (h and i) Integrated emission spectrum for emitter 2 (at 54-55 and 1369-1370 s) and histogram of emission maxima (j). Distribution of photon counts (k). (l) Histogram showing average number of frames for 208 emitters. Distributions of emission maxima measured for emitters that display 'low' blinking (less than 100 on-off cycles and median 'off' time greater than 30 s) (m) and 'high' (greater than 600 on-off cycles and median 'off' time greater than 30 s) (n).

# Supplementary Information.

# Photodynamics of single colour emitters in large-area monolayers of hexagonal-boron nitride.

H. L. Stern*[1] and R. Wang[2], Y. Fan[2], R, Mizuta[2], J.C. Stewart[2], L.M. Needham[1], T. D, Roberts[3], R. Wai[3], N. S, Ginsberg[3], D. Klenerman[1], S. Hofmann*[2] and S. F. Lee[1].

[1] Department of Chemistry, University of Cambridge, Lensfield Road, CB2 1EW, Cambridge, United Kingdom.

[2] Department of Engineering, University of Cambridge, JJ Thompson Avenue, CB3 0FA, Cambridge, United Kingdom.

[3] Department of Chemistry, University of California, Berkeley, CA 94720.

**Figure S1. Transmission electron microscopy (TEM) images of the h-BN monolayer.**

(a) TEM map of crystalline domains in an h-BN monolayer. The map is coloured by the crystal orientation measured by micro region diffraction. (b) TEM diffraction pattern measured for an h-BN monolayer. (c) Histogram of crystal orientation.

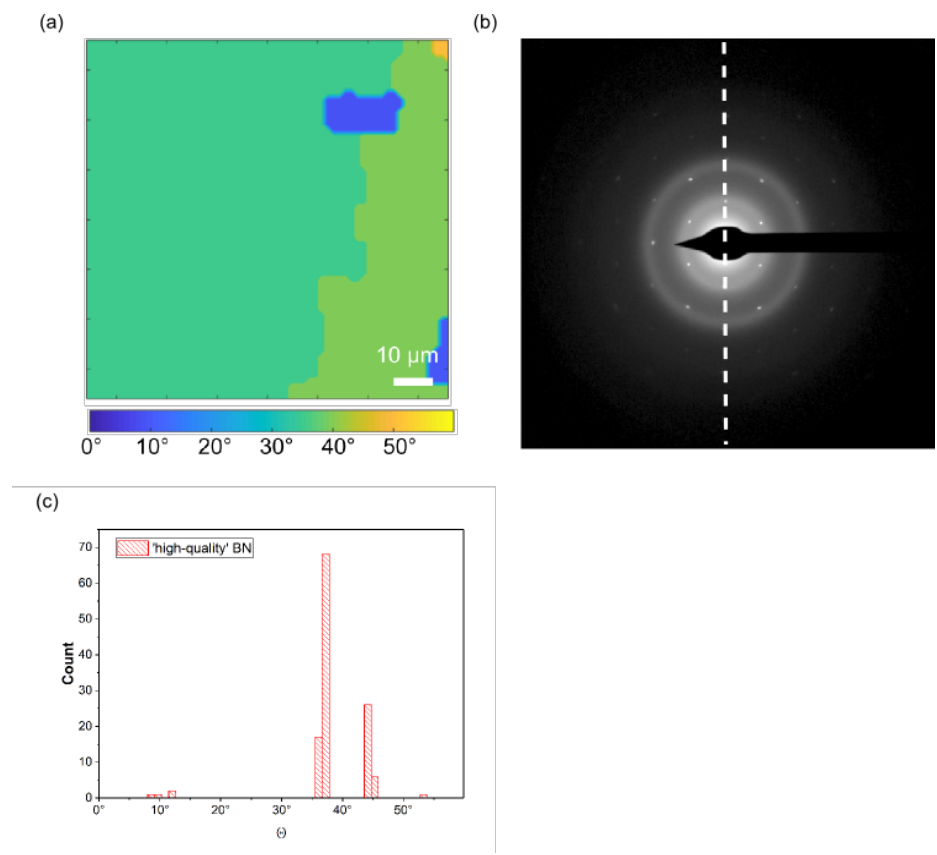

**Transmission grating spectral calibration.**

Spectral calibration of the camera with the transmission grating is done via imaging of TetraSpeck™ beads at three different excitation wavelengths: 405 nm, 532 nm and 633 nm. Excitation of the beads at each of these wavelengths results in a well-documented emission spectrum with defined peak positions (centred at 512.7 nm, 581.5 nm, and 676.5 nm). The emission peaks are generated on the spectral domain area of the chip. The intensities over all localisations are summed, and the centre positions (in pixels) of the peaks are determined by fitting to three different Gaussian distributions. The diffraction distance for a given each of the three emission wavelengths is plotted against the x and y pixel coordinates to obtain a relationship between nm/pixel ratio across the camera area.

In Figure S2 we demonstrate the fitting procedure with data from the 532 nm calibration.

**Figure S2:** Tetraspeck[TM] beads excited with 532 nm. (a) An integrated fluorescence plot along the x axis of the camera, for a single bead, showing the spatial domain and spectral domain of the camera. The emission spectrum of the beads when excited at 532 nm shows an emission maximum at 581.5 nm. (b) Example of four bead spectra (blue dashed line) extracted from the spectral part of the image (Tetraspeck[TM] beads excited with 532 nm). For each spectrum a 20 pixel region around the maximum is fit to a Gaussian distribution (red line) to determine the centre position.

(a)

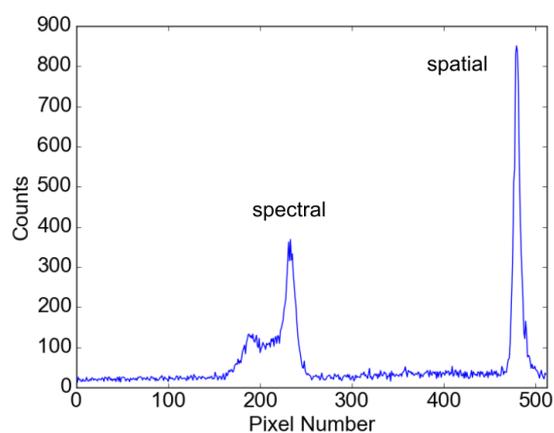

(b)

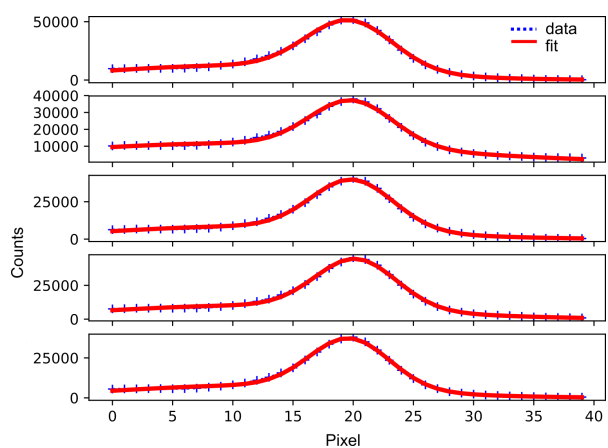

**Figure S3: Calibration of camera area.** By exciting TetraSpeck™ beads at 405 nm, 532 nm and 633 nm, we obtain the diffraction distances for the emissive wavelengths 512.7 nm, 581.5 nm, and 676.5 nm on the spectral region of the camera, as a function of x,y bead position in the spatial region of the camera. The image below shows the relationship between x and y coordinates and wavelength.

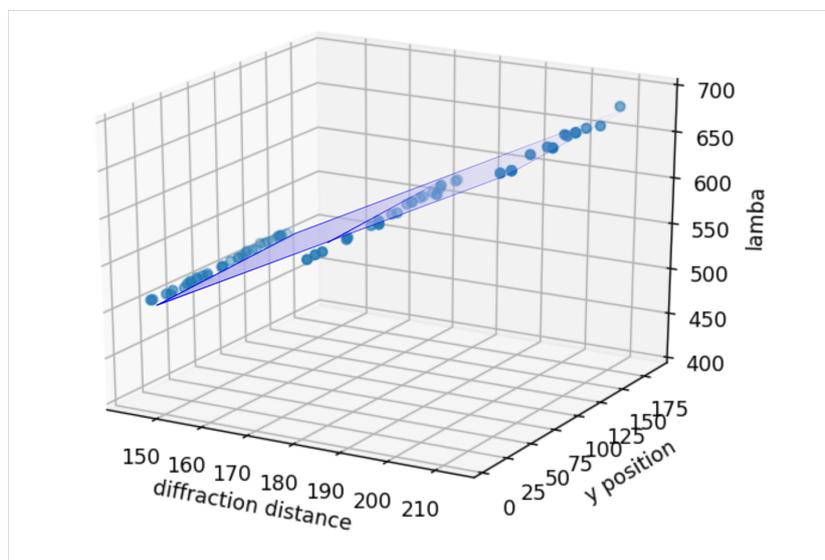

**Figure S4: Spatial and spectral localisation precision of the spectrally-resolved set-up.**

Dependence of the experimentally-obtained spatial localisation precision on the measured peak photon signal(a). Corresponding spectral localization precision of the setup (b), as per reference 20. Briefly, the deviation in the mean centre position of a spatially isolated and immobile bead was determined as a function of the photon emission and plotted below.

In the measurements of h-BN, we analysed localisations with spatial precision of 10-30 nm, corresponding to peak detected photon signals of ~100-1000 per 100 ms frame and a spectral resolution of 2-8 nm. This corresponds to an integrated photon count of typically $1.0 \times 10^4 - 3.5 \times 10^4$ photons/s. This thresh-holding was chosen in order to have the best compromise between spatial precision and resolution in the spectral domain.

(a)

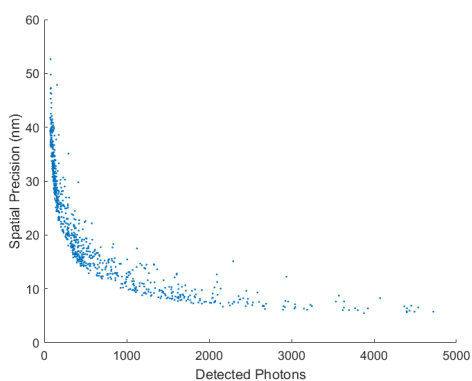 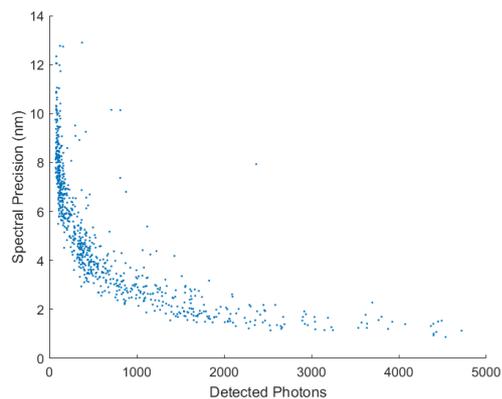

(b)

**Figure S5: Spatial Precision of localisations**

Histogram of the distribution of spatial precision obtained from single emitters of h-BN. We selected localisations with a spatial precision of 30 nm and under, corresponding to a spectral precision of ~ 8 nm and under.

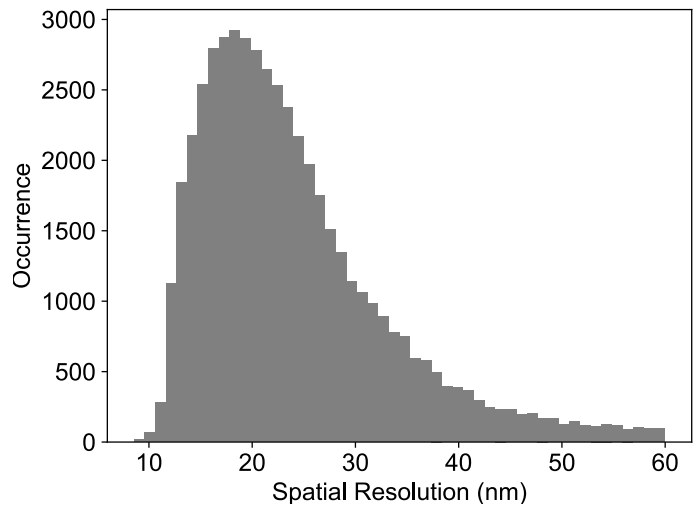

**Figure S6: Distribution of peak photons detected**

Peak photons detected for the h-BN emitters range from 100-800 photons per 100 ms frame. This corresponds to an integrated photon count of typically 1.0-3.5 x $10^4$ counts/s.

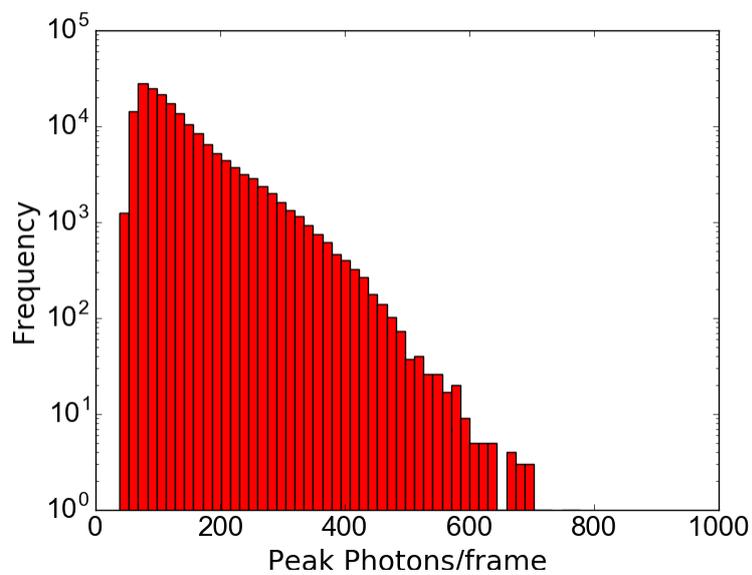

**Figure S7: Photobleaching.**

To understand the global photobleaching rate of the emitters, we compared ten measurements from three samples (red, green and blue). The total number of detected photons per frame was compiled and plotted against frame number, normalized to the photon count at the start of the measurement. We compared the decay in photons detected to a blank slide with organic residue (grey). We find that organic contaminant has bleached ~80% by 250 frames and fully bleached within ~500 frames (50 s). The h-BN emitters have lost 40% of brightness by 2000 frames (200 s).

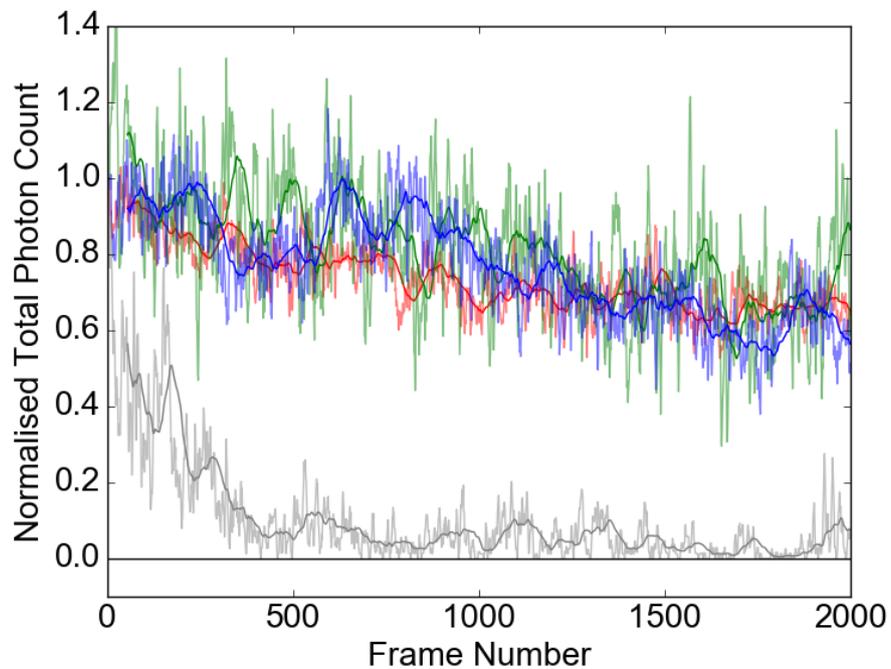

**Figure S8: Spectral characterization of organic contaminant.**

To determine the spectral profile of organic contaminant from the transfer process, we measured ten blanks slides that had been exposed to the same transfer process and not annealed. These measurements were processed in the same way as the measurements of h-BN.

Over 10 measurements we observe 10% fewer emitters on the blank slide, than we observe in ten h-BN measurements (red trace). The contamination on the slides emits light predominantly at 610-620 nm.

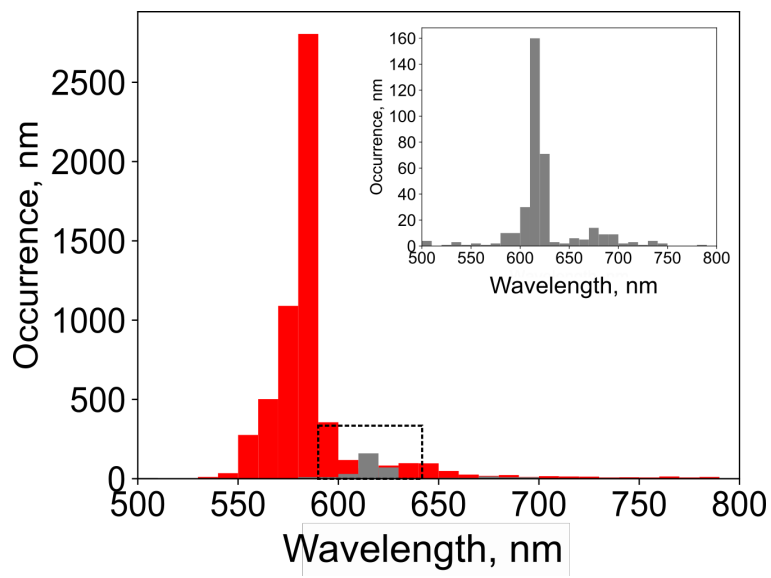

**Figure S9. Single Emitter Spectrum**

Representative example of a typical emission spectrum collected for an emitter, from one 100 ms frame, background subtracted.

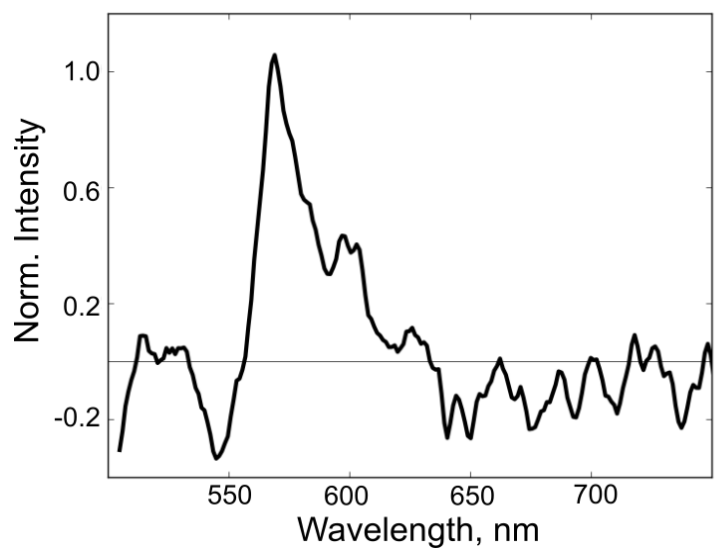